\documentclass
[floats,amsmath,amssymb,4paper,notitlepage,onecolumn,aps]{revtex4}%
\usepackage{bm}
\usepackage{amsmath}
\usepackage{amsfonts}
\usepackage{amssymb}
\usepackage{graphicx}%
\providecommand{\U}[1]{\protect\rule{.1in}{.1in}}
\pdfoutput=1
\begin{document}
\title{Spin Caloritronics}
\author{Gerrit E. W. Bauer$^{1,2}$}
\affiliation{$^{1}$Institute for Materials Research, Tohoku University, 2-1-1 Katahira,
Aoba-ku, Sendai 980-8577, Japan}
\affiliation{$^{2}$Delft University of Technology, Kavli Institute of NanoScience,
Lorentzweg 1, 2628 CJ Delft, The Netherlands}

\begin{abstract}
This is a brief overview of the state of the art of spin caloritronics, the
science and technology of controlling heat currents by the electron spin
degree of freedom (and \textit{vice versa}). 

\end{abstract}
\maketitle

\section{Introduction}

The coupling between spin and charge transport in condensed matter is studied
in the lively field referred to as spintronics. Heat currents are coupled to
both charge and spin currents \cite{Johnson87,Johnson10}. `Spin caloritronics'
is the field combining thermoelectrics with spintronics and nanomagnetism,
which recently enjoys renewed attention \cite{Bauer10a}. The term
\textquotedblleft caloritronics\textquotedblright\ (from `calor', the Latin
word for heat) has recently been introduced to describe the endeavor to
control heat transport on micro- and nanometer scales. Alternative expressions
such as \textquotedblleft(mesoscopic) heattronics\textquotedblright\ or
\textquotedblleft caloric transport\textquotedblright\ have also been
suggested. Specifically, spin caloritronics is concerned with new physics
related to spin, charge and entropy/energy transport in materials and
nanoscale structures and devices. Examples are spin dependence of thermal
conductance, Seebeck and Peltier effects, heat current effects on spin
transfer torque, thermal spin and anomalous Hall effects, \textit{etc}. Heat
and spin effects are also coupled by the dissipation and noise associated with
magnetization dynamics.

The societal relevance of the topic is given by the imminent breakdown of
Moore's Law by the thermodynamic bottleneck: further decrease in feature size
and transistor speed goes in parallel with intolerable levels of Ohmic energy
dissipation associated with the motion of electrons in conducting circuits.
Thermoelectric effects in meso- \cite{Giazotto} and nanoscopic \cite{Dubi}
structures might help in managing the generated heat. Spin caloritronics is
intimately related to possible solutions to these problems by making use of
the electron spin degree of freedom.

Spin caloritronics is as old as spin electronics, starting in the late 1980's
with M. Johnson and R.H. Silsbee's \cite{Johnson87} visionary theoretical
insights into the non-equilibrium thermodynamics of spin, charge and heat in
metallic heterostructures with collinear magnetization configurations. Except
for a few experimental studies on the thermoelectric properties of magnetic
multilayers in the CIP (currents in the interface plane) configuration
\cite{Shi96} in the wake of the discovery of the giant magnetoresistance, the
field remained dormant for many years. The Lausanne group started systematic
experimental work on what we now call spin caloritronics in magnetic
multilayer nanowires and further developed the theory \cite{Gravier06}.

Several new and partly unpublished discoveries in the field of spin
caloritronics excite the community, such as the spin (wave) Seebeck effect in
and signal transmission through magnetic insulators, the spin-dependent
Seebeck effect in magnetic nanostructures, the magnonic thermal Hall effect,
giant Peltier effect in constantan/gold nanopillars, and the thermal spin
transfer torque. After a brief introduction into the basics of how the spin
affects classical thermoelectric phenomena, these topics will appear in the
following sections.

\section{Basic physics}

We learn from textbooks that the electron-hole asymmetry at the Fermi energy
in metals generates thermoelectric phenomena. A heat current $\mathbf{\dot{Q}%
}$ then drags charges with it, thereby generating a thermopower voltage or
charge current $\mathbf{J}$ for open or closed circuit conditions,
respectively. \textit{Vice versa} a charge current is associated by a heat
current, which can be used to heat or cool the reservoirs. In a diffusive bulk
metal the relation between the\ local driving forces, \textit{i.e}. the
voltage gradient or electric field $\mathbf{E}=\boldsymbol{\nabla}%
_{\mathbf{r}}V$ and temperature gradient $\boldsymbol{\nabla}_{\mathbf{r}}T$
reads
\begin{equation}
\left(
\begin{array}
[c]{c}%
\mathbf{J}\\
\mathbf{\dot{Q}}%
\end{array}
\right)  =\sigma\left(
\begin{array}
[c]{cc}%
1 & S\\
\Pi & \kappa/\sigma
\end{array}
\right)  \left(
\begin{array}
[c]{c}%
\boldsymbol{\nabla}_{\mathbf{r}}V\\
-\boldsymbol{\nabla}_{\mathbf{r}}T
\end{array}
\right)  . \label{TE}%
\end{equation}
where $\sigma$ is the electric conductivity, $S$ the Seebeck coefficient and
$\kappa$ the heat conductivity \cite{Ashcroft}. The Kelvin-Onsager relation
between the Seebeck and Peltier coefficients $\Pi=$ $ST$ is a consequence of
Onsager reciprocity \cite{Onsager}. In the Sommerfeld approximation, valid
when the conductivity as a function of energy varies linearly on the scale of
the thermal energy $k_{B}T$ or, more precisely, when $\mathcal{L}_{0}%
T^{2}\left\vert \partial_{\varepsilon}^{2}\sigma\left(  \varepsilon\right)
|_{\varepsilon_{F}}\right\vert \ll\sigma\left(  \varepsilon_{F}\right)  $,
\begin{equation}
S=-e\mathcal{L}_{0}T\frac{\partial}{\partial\varepsilon}\ln\sigma
(\varepsilon)|_{\varepsilon_{F}},
\end{equation}
where the Lorenz constant $\mathcal{L}_{0}=\left(  \pi^{2}/3\right)  \left(
k_{B}/e\right)  ^{2}$ and $\sigma\left(  \varepsilon\right)  $ is the
energy-dependent conductivity around the Fermi energy $\varepsilon_{F}$. In
this regime the Wiedemann-Franz Law
\begin{equation}
\kappa=\sigma\mathcal{L}_{0}T
\end{equation}
holds. Thermoelectric phenomena at constrictions and interfaces are obtained
by replacing the gradients by differences and the conductivities by conductances.

The spin dependence of the thermoelectric properties in isotropic and
monodomain metallic ferromagnets can be expressed in the two-current model of
majority and minority spins \cite{Johnson87,Gravier06,Hatami09,Tatara}:
\begin{equation}
\left(
\begin{array}
[c]{c}%
\mathbf{J}_{c}\\
\mathbf{J}_{s}\\
\mathbf{\dot{Q}}%
\end{array}
\right)  =\sigma\left(
\begin{array}
[c]{ccc}%
1 & P & S\\
P & 1 & P^{\prime}S\\
ST & P^{\prime}ST & \mathcal{L}_{0}T
\end{array}
\right)  \left(
\begin{array}
[c]{c}%
\boldsymbol{\nabla}_{\mathbf{r}}\tilde{\mu}_{c}/e\\
\boldsymbol{\nabla}_{\mathbf{r}}\mu_{s}/2e\\
-\boldsymbol{\nabla}_{\mathbf{r}}T
\end{array}
\right)  , \label{ons1}%
\end{equation}
where $\mathbf{J}_{c(s)}=\mathbf{J}^{\left(  \uparrow\right)  }\pm
\mathbf{J}^{\left(  \downarrow\right)  }$ and $\mathbf{\dot{Q}}=\mathbf{\dot
{Q}}^{\left(  \uparrow\right)  }+\mathbf{\dot{Q}}^{\left(  \downarrow\right)
}$ are the charge, spin and heat currents, respectively. $P$ and $P^{\prime}$
stand for the spin-polarization of the conductivity and its energy derivative%
\begin{equation}
P=\left.  \frac{\sigma^{\left(  \uparrow\right)  }-\sigma^{\left(
\uparrow\right)  }}{\sigma^{\left(  \uparrow\right)  }+\sigma^{\left(
\uparrow\right)  }}\right\vert _{\varepsilon_{F}};\;P^{\prime}=\left.
\frac{\partial_{\varepsilon}\sigma^{\left(  \uparrow\right)  }-\partial
_{\varepsilon}\sigma^{\left(  \uparrow\right)  }}{\partial_{\varepsilon}%
\sigma^{\left(  \uparrow\right)  }+\partial_{\varepsilon}\sigma^{\left(
\uparrow\right)  }}\right\vert _{\varepsilon_{F}}.
\end{equation}
$\tilde{\mu}_{c}=\left(  \mu^{\left(  \uparrow\right)  }+\mu^{\left(
\downarrow\right)  }\right)  /2\ $is\ the charge electrochemical potential and
$\mu_{s}=\mu^{\left(  \uparrow\right)  }-\mu^{\left(  \downarrow\right)  }$
the difference between chemical potentials of the two-spin species,
\textit{i.e.} the spin accumulation. The spin-dependent thermal conductivities
obey the Wiedemann-Franz law $\kappa^{\left(  \alpha\right)  }\approx
\mathcal{L}_{0}T\sigma^{\left(  \alpha\right)  }$ when $S^{\uparrow
(\downarrow)}\ll\sqrt{\mathcal{L}_{0}}$ and the total thermal conductivity
$\kappa=\kappa^{\left(  \uparrow\right)  }+\kappa^{\downarrow}=\mathcal{L}%
_{0}T\sigma$. In Eq. (\ref{ons1}) the spin heat current $\mathbf{\dot{Q}}%
_{s}=\mathbf{\dot{Q}}^{\left(  \uparrow\right)  }-\mathbf{\dot{Q}}^{\left(
\downarrow\right)  }$ does not appear. This is a consequence of the implicit
assumption that there is no spin temperature (gradient) $T_{s}=T^{\left(
\uparrow\right)  }-T^{\left(  \downarrow\right)  }\ $due to effective
interspin and electron-phonon scattering \cite{Hatami09}. This approximation
does not necessarily hold at the nanoscale and low temperatures
\cite{Heiki1,Heiki2}. Although initial experiments were inconclusive, a
lateral spin valve device has been proposed in which it should be possible to
detect spin temperatures.

Above equations presume that the spin projections are good quantum numbers,
which is not the case in the presences of non-collinear magnetizations or
significant spin-orbit interactions. Both complications give rise to new
physics in spintronics, such as the spin Hall effect and current-induced spin
transfer torques. Both have their spin caloritronic equivalents.

Lattice vibrations (phonons) provide a parallel channel for heat currents, as,
in magnets, do spin waves (magnons). The study and control of spin waves is
referred to as `Magnonics' \cite{Magnonics}. The coupling of different modes
can be very important for thermoelectric phenomena, causing for instance the
phonon-drag effect on the thermopower at lower temperatures. The heat current
carried by magnons is a spin current and may affect the Seebeck coefficient
\cite{Tular}. In metallic ferromagnets the spin wave heat current appears to
be smaller than the thermoelectric heat current discussed above, but is the
dominant mode of spin transport in magnetic insulators \cite{Hess,Meier}. The
coupling between magnons and phonons has been recently demonstrated in the
spin Seebeck effect (see Sec. \ref{SSE} and the Chapter by E. Saitoh).

\section{Spin-dependent thermoelectric phenomena in metallic structures}

A consequence of the basics physics sketched above is the existence of
thermoelectric generalizations of the giant magnetoresistance (GMR),\textit{
i.e.} the modulation of the electric charge and heat currents by the spin
configuration of magnetic multilayers, spin valves and tunneling junctions as
well as a family of thermal spin Hall effects.

\subsection{Magneto-Peltier and Seebeck effects\label{SDSE}}

The magneto-Peltier and magneto-Seebeck effects are caused by the
spin-dependence of the Seebeck/Peltier coefficients in ferromagnets
\cite{Johnson87,Gravier06,Hatami09}. The magnetothermopower has been observed
in multilayered magnetic nanowires \cite{Gravier06}. A large Peltier effect in
constantan (CuNi alloy)/Au \cite{Fukushima} has been associated with magnetism
in the phase-separation magnetic phase \cite{Yoshida}.

A magneto-Seebeck effect in lateral spin valves has been demonstrated
\cite{Slachter}. Here a temperature gradient is intentionally applied over an
intermetallic interface. The spin-dependence of the Seebeck coefficient induce
a spin-polarized current into the normal metal, in which Slachter \textit{et
al.} \cite{Slachter} detect the accompanying spin accumulation by an analyzing
ferromagnetic contact. A spin-dependent thermopower has been predicted for
molecular spin valves from first-principles theory \cite{Mertig}. A magneto
Seebeck effect in magnetic tunnel junctions has been observed
\cite{Schumacher,MTJ} and modelled by ab initio calculations \cite{Heiliger}.
A spin-dependent Seebeck effect in Py$|$Si tunneling junctions has been
observed by Le Breton \textit{et al}. \cite{Ronnie} by analyzing the magnetic
field dephasing (Hanle effect) of a thermally injected spin accumulation. The
thermoelectric figure of merit can possibly be improved by employing the
conducting edge and surface states of topological insulators \cite{Murakami}.

\subsection{Thermal Hall effects}

\begin{figure}[ptb]%
\centering
\includegraphics[
natheight=4.568000in,
natwidth=11.296400in,
height=1.0172in,
width=2.5025in
]%
{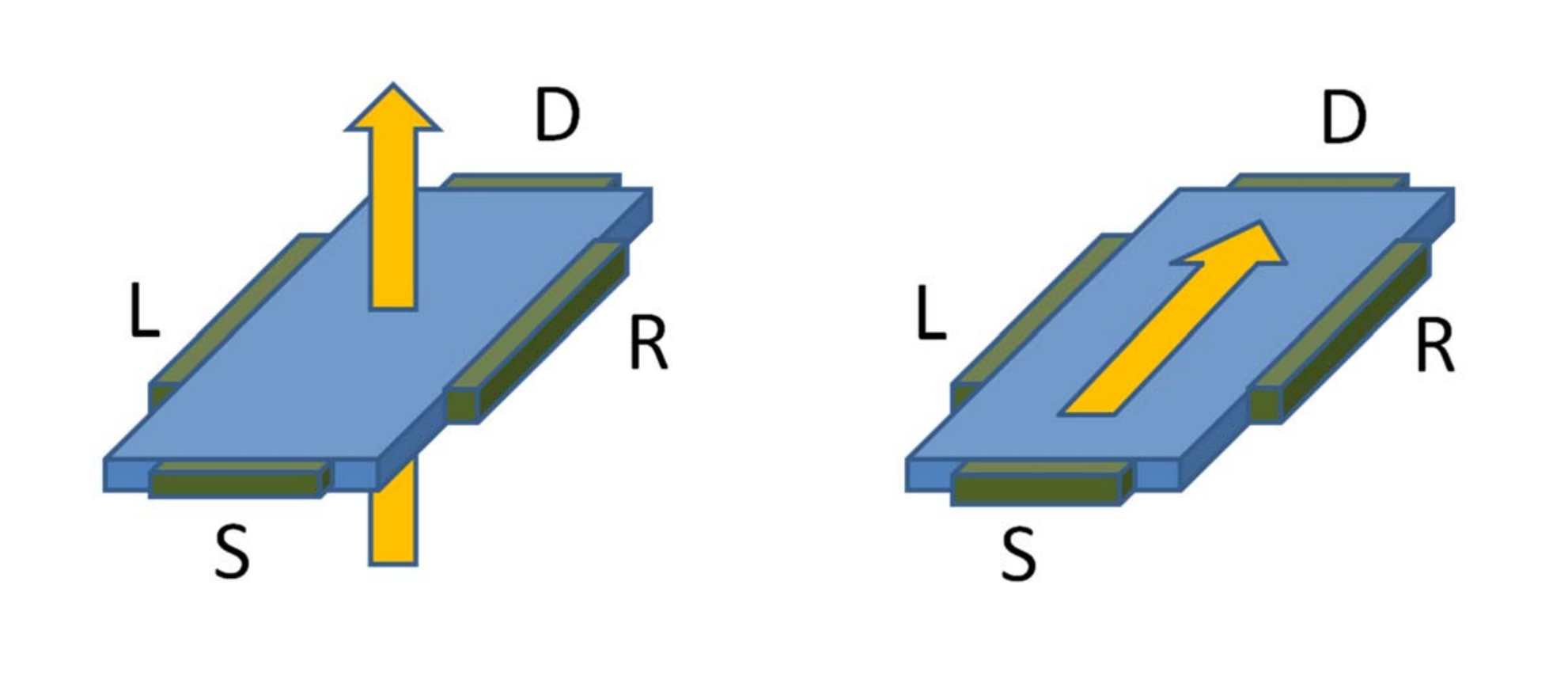}%
\caption{A sketch of the configuration of anomalous (left figure) and planar
(right figure) Hall effects in ferromagnets. S and D denote source and drain
contacts and L and R left and right Hall contacts. The arrow denotes the
magnetization direction.}%
\label{Hall}%
\end{figure}

Thermal Hall effects exist in normal metals in the presence of external
magnetic fields and can be classified into three groups \cite{Callen}. The
Nernst effect stands for the Hall voltage induced by a heat current. The
Nettingshausen effect describes the heat current induced transverse to an
applied charge current. The Hall heat current induced by a temperature
gradient goes by the name of Righi-Leduc. The spin degree of freedom opens a
family of spin caloritronic Hall effects in the absence of an external field
which are not yet fully explored. We may add the label spin in order to
describe effects in normal metals (spin Hall effect, spin Nernst effect,
\textit{etc}.). In ferromagnets we may distinguish the configuration in which
the magnetization is normal to both currents (anomalous Hall effect, anomalous
Nernst effect, \textit{etc}.) from the configuration with in-plane
magnetization (planar Hall effect, planar Nernst effect, \textit{etc}.) as
sketched in Figure \ref{Hall}. Theoretical work has been carried out with
emphasis on the intrinsic spin-orbit interaction \cite{Onoda,Ma,Xie}.

Seki \textit{et al.} \cite{Seki} found experimental evidence for a thermal
Hall effect in Au$|$FePt structures, which can be due either to an anomalous
Nernst effect in FePt or a spin Nernst effect in Au. In GaMnAs the planar
\cite{Pu1} and anomalous \cite{Pu2} Nernst effects have been observed, with
intriguing temperature dependences. Slachter \textit{et al}. \cite{Slachter11}
identified the anomalous Nernst effect and anisotropic magnetoheating in
multiterminal permalloy$|$copper spin valves.%

\section{Thermal spin transfer torques}

A spin current is in general not conserved. In a metal, angular momentum can
be dissipated to the lattice by spin-flip scattering. In the presence of a
non-collinear magnetic texture, either in a heterostructure, such as a spin
valve and tunnel junction, or a magnetization texture such as a domain wall or
magnetic vortex, the magnetic condensate also absorbs a spin current, which by
conservation of angular momentum leads to a torque on the magnetization that,
if strong enough, can lead to coherent magnetization precessions and even
magnetization reversal \cite{Ralph}. Just like a charge current, a heat
current can exert a torque on the magnetization as well \cite{Hatami07}, which
leads to purely thermally induced magnetization dynamics \cite{Wegrowe}. Such
a torque can be measured under closed circuit conditions, in which part of the
torque is simply exerted by the spin-dependent thermopower, and in an open
circuit in which a charge current is suppressed \cite{Hatami07}.

\subsection{Spin valves \label{SV}}

The angular dependence of the thermal torque can be computed by circuit theory
\cite{Hatami07,Hatami09}. Thermal spin transfer torques have been detected in
nanowire spin valves \cite{Yu}. Slonczewski \cite{Slonc} studied the spin
transfer torque in spin valves in which the polarizer is a magnetic insulator
that exerts a torque on a free magnetic layer in the presence of a temperature
gradient. He concludes that the thermal torque can be more effective in
switching magnetizations than a charge current-induced torque. Note that the
physics of heat current induced spin injection by magnetic insulators is
identical to that of the longitudinal spin Seebeck effect as discussed briefly
in Sec. \ref{SSE}.

\subsection{Magnetic tunnel junctions}

Large thermal torques have been predicted by first-principles calculations for
magnetic tunnel junctions with thin barriers that compare favorable with those
obtainable by an electric bias \cite{Jia1}, but these have as yet not been
confirmed experimentally.

\subsection{Textures}

Charge current-induced magnetization in magnetic textures have enjoyed a lot
of attention in recent years. Domain wall motion can be understood easily in
terms of angular momentum conservation in the adiabatic regime, in which the
length scale of the magnetization texture such as the domain wall width is
much larger than the scattering mean free path or Fermi wave length, as
appropriate for most transition metal ferromagnets. In spite of initial
controversies, the importance of dissipation in the adiabatic regime
\cite{Zhang} is now generally appreciated. In analogy to the Gilbert damping
factor $\alpha$ the dissipation under an applied current is governed by a
material parameter $\beta_{c}$ that for itinerant magnetic materials is of the
same order as $\alpha$ \cite{Tserkovnyak}. In the case of a heat-current
induced domain wall motion, the adiabatic thermal spin transfer torque
\cite{Hatami07} is also associated with a dissipative $\beta_{T}$-factor that
is independent of the charge-current $\beta_{c}$ \cite{Kovalev09,Bauer10}.
$\beta_{T}$ has been explicitly calculated by Hals for GaMnAs \cite{Hals}.
Non-adiabatic corrections to the thermal spin transfer torque in fast-pitch
ballistic domain walls have been calculated by first-principles \cite{Yuan}.
Laser induced domain wall pinning might give clues for heat current effects on
domain wall motion \cite{Klaui}.

In insulating ferromagnets, domain wall still be moved since part of the heat
current is carried by spin waves, and therefore associated with angular
momentum currents. In contrast to metals in which the angular momentum current
can have either sign relative to the heat current direction, in insulators the
magnetization current flows always against the heat current, which means that
the adiabatic torque moves the domain wall to the hot region
\cite{Nowak,Kovalev2,XR}.

\section{Magneto-heat resistance}

The heat conductance of spin valves is expected to depend on the magnetic
configuration, similar to the GMR, giving rise to a giant magneto-heat
resistance \cite{Hatami07} or a magnetotunneling heat resistance. In contrast
to the GMR, the magnetoheat resistance is very sensitive to inelastic
(interspin and electron-phonon) scattering \cite{Heiki1,Heiki2}.

\begin{figure}[ptb]%
\centering
\includegraphics[
trim=0.054768in 0.000000in -0.054768in 0.000000in,
natheight=5.173800in,
natwidth=5.617200in,
height=2.4505in,
width=2.6585in
]%
{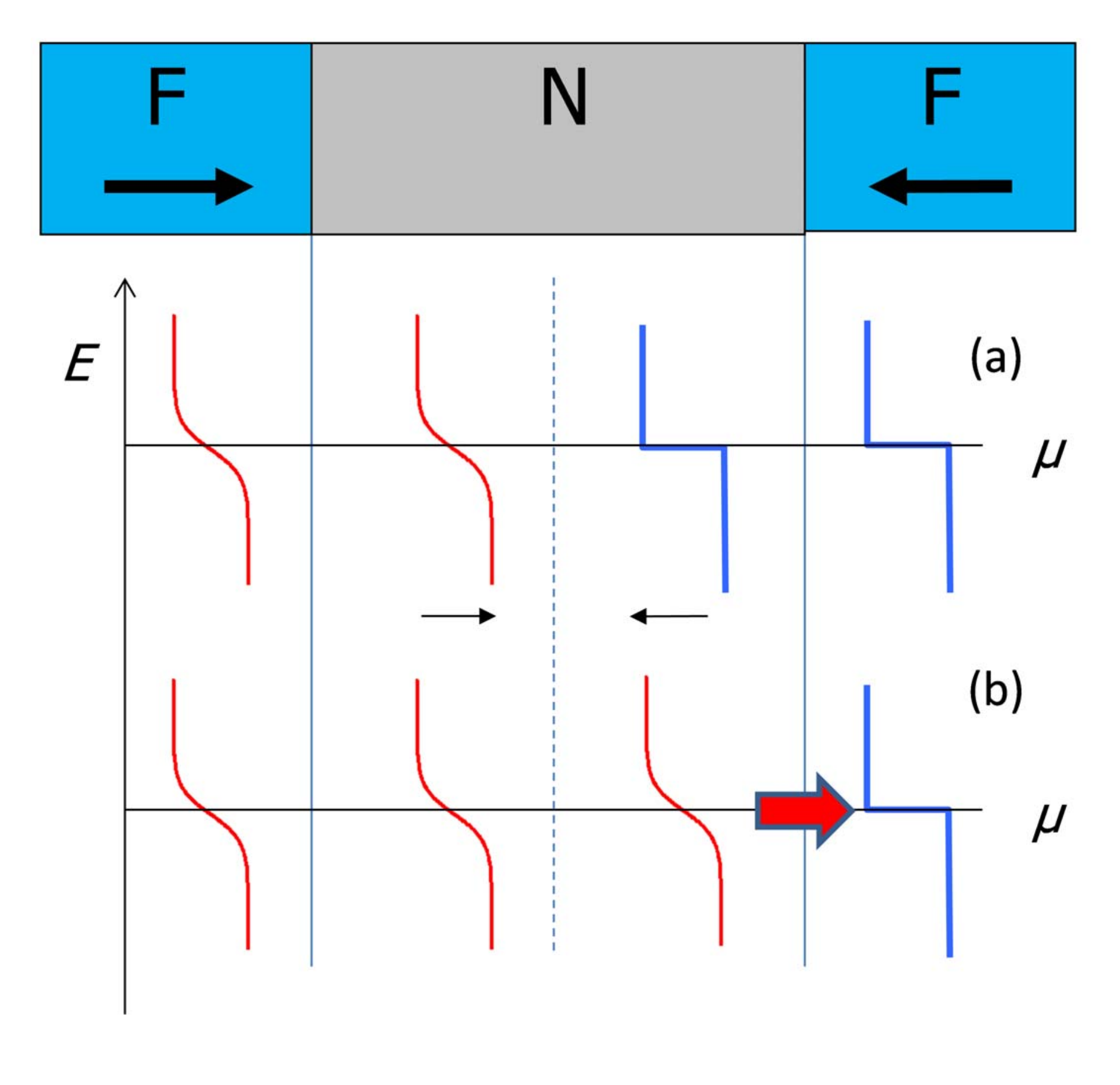}%
\caption{A temperature difference over a spin valve with half-metallic
contacts and an antiparallel configuration of the magnetic contacts. Plotted
are the electron distribution functions in the ferromagnets and the normal
metal spacer ($\mu$ is the chemical potential). In (a) the spins in the spacer
are non-interacting, in (b) they are strongly interacting, thereby allowing a
heat current flow through the left interface.}%
\label{WF}%
\end{figure}

Inelastic scattering leads to a breakdown of the Wiedemann-Franz Law in spin
valves. This is most easily demonstrated for half-metallic ferromagnetic
contacts as sketched in Fig. \ref{WF} for a finite temperature bias over the
sample. In the figure the distribution functions are sketched for the three
spatial regions. Both spins form eigenstates in N, but in F only the majority
spin exists. In Fig. \ref{WF}(a) we suppose absence of inelastic scattering
between the spins, either by direct Coulomb interaction or indirect energy
exchange via the phonons. When a strong interaction is switched on both spins
in N will adopt the same temperature as sketched in Fig. \ref{WF}(b). The
temperature gradient on the right interface will induce a heat current, while
a charge current is suppressed, clearly violating the Wiedemann-Franz Law. A
spin heat valve effect can therefore only exist when the interspin and
spin-phonon interactions are sufficiently weak.

\begin{figure}[ptb]%
\centering
\includegraphics[
natheight=11.113900in,
natwidth=11.691300in,
height=1.6303in,
width=1.7143in
]%
{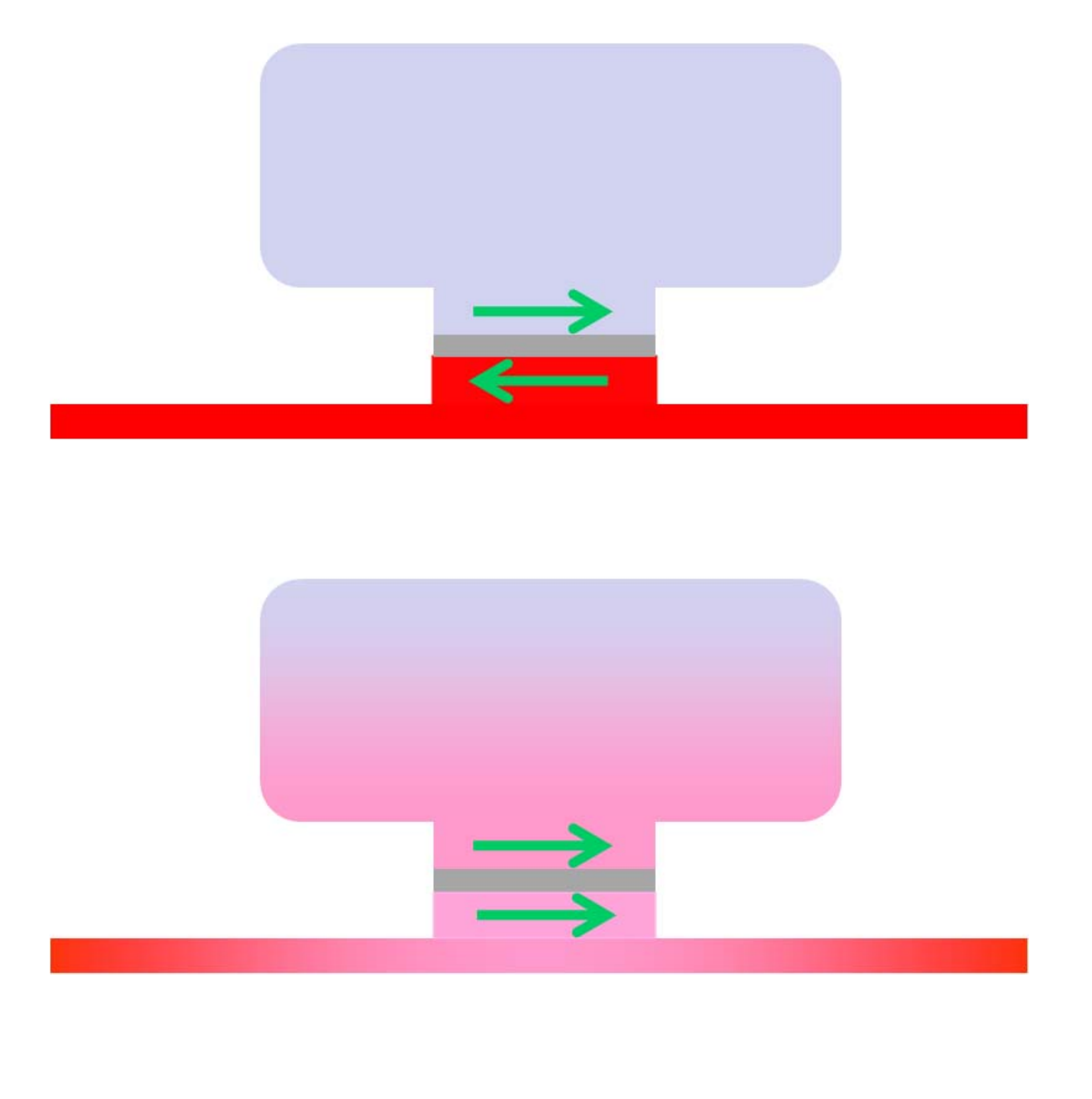}%
\caption{The dependence of the heat conductance of a magnetic tunnel junction
or spin valve on the magnetic configuration can be used to control possible
overheating of a substrate, such as a hot spot in an integrated circuit, when
the necessity arises.}%
\label{HS}%
\end{figure}

The heat conductance of tunnel junctions is expected to be less sensitive to
inelastic scattering. A useful application for on-chip heat management could
be a tunneling heat valve, \textit{i.e}. a switchable heat sink as illustrated
in Fig. \ref{HS}.%

\section{Spin caloritronic heat engines and motors}

\begin{figure}[ptb]%
\centering
\includegraphics[
natheight=3.934900in,
natwidth=10.135000in,
height=2.0965in,
width=5.3681in
]%
{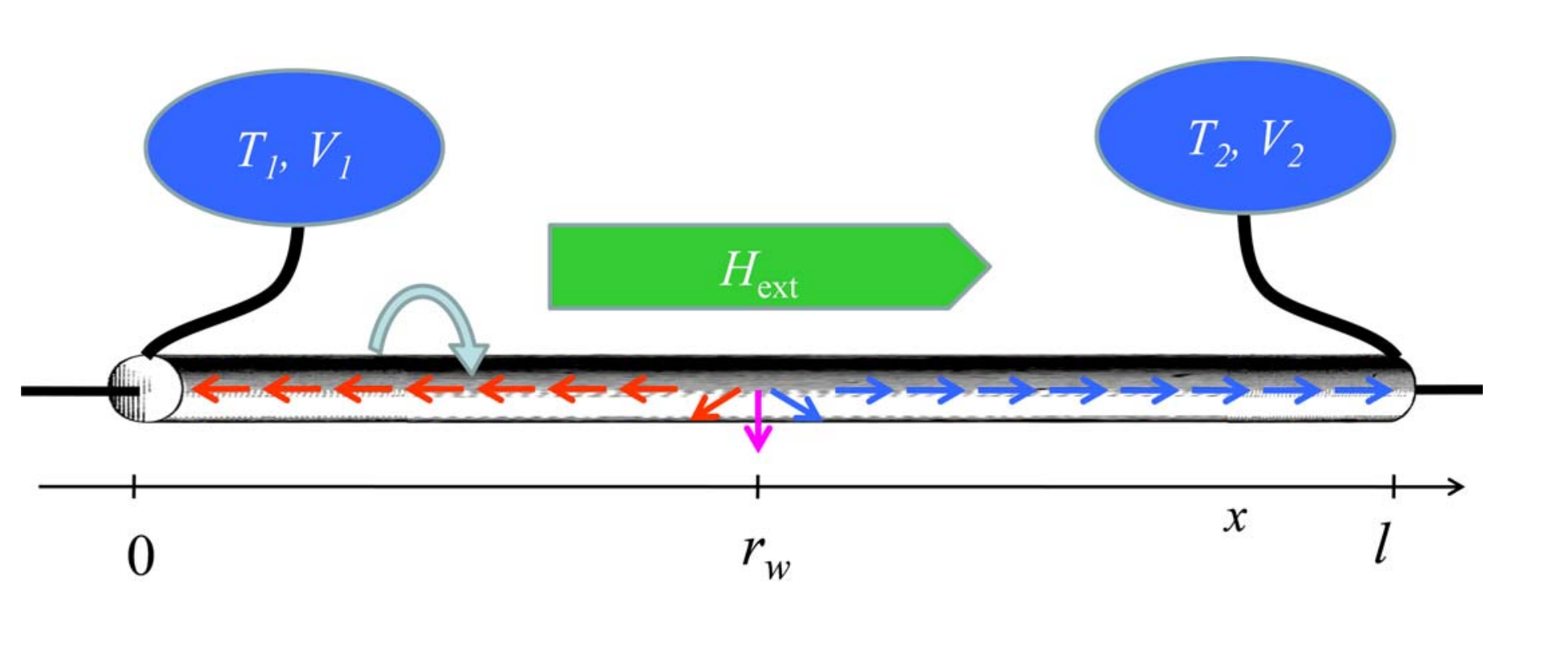}%
\caption{Magnetic nanowire of length $l$ in electrical and thermal contact
with reservoirs. A domain wall is centered at position $r_{w}$. The wire is
mounted such that it can rotate around the $x$-axis. A magnetic field and
mechanical torque can be applied along $x.$}%
\label{Motor}%
\end{figure}

Onsager's reciprocal relations \cite{Onsager} reveal that seemingly unrelated
phenomena can be expressions of identical microscopic correlations between
thermodynamic variables of a given system \cite{Groot}. The archetypal example
is the Onsager-Kelvin identity of thermopower and Peltier cooling mentioned
earlier. We have seen that spin and charge currents are coupled with each
other and with the magnetization. Furthermore, mechanical and magnetic
excitations are coupled by the Barnett and Einstein-de Haas effects
\cite{Barnett,EinsteindeHaas}. The thermoelectric response matrix including
all these variables can be readily formulated for a simple model system
consisting of a rotatable magnetic wire including a domain wall as sketched in
Fig. \ref{Motor}.%
The linear response matrix then reads $\mathbf{J}=\hat{L}\mathbf{X}$, where
the generalized currents $\mathbf{J}$ and forces $\mathbf{X}$
\begin{align}
\mathbf{J}  &  =\left(
\begin{array}
[c]{cccc}%
J_{c}, & J_{Q}, & \dot{\varphi}, & \dot{r}_{w}%
\end{array}
\right)  ^{T}\\
\mathbf{X}  &  =\left(
\begin{array}
[c]{cccc}%
-\Delta V, & -\frac{\Delta T}{T}, & \tau_{\mathrm{ext}}^{\mathrm{mech}}, &
-2AM_{s}H_{\mathrm{ext}}%
\end{array}
\right)  ^{T}%
\end{align}
are related by the response matrix
\begin{equation}
\hat{L}=\left(
\begin{array}
[c]{cccc}%
L_{cc} & L_{cQ} & L_{c\varphi} & L_{cw}\\
L_{Qc} & L_{QQ} & L_{Q\varphi} & L_{Qw}\\
L_{\varphi c} & L_{\varphi Q} & L_{\varphi\varphi} & L_{\varphi w}\\
L_{wc} & L_{wQ} & L_{w\varphi} & L_{ww}%
\end{array}
\right)  . \label{Onsager}%
\end{equation}
Onsager reciprocity implies that $L_{xy}=\pm L_{yx}$. The elements can be
computed by scattering theory \cite{Bauer10}.

The matrix relation between generalized forces and currents implies a large
functionality of magnetic materials. Each of the forces can give rise to all
currents, where a temperature gradient is especially relevant here. The
response coefficient $L_{cQ}$ clearly represents the Seebeck effect, $L_{QQ}$
the heat conductance, $L_{\varphi Q}$ a thermally driven (Brownian) motor, and
$L_{wQ}$ a heat current driven domain wall motion \cite{Kovalev09}. Onsager
symmetry implies that $L_{wQ}=L_{Qw}$ and $L_{\varphi Q}=-L_{Q\varphi}$.
\textit{E.g}. a Peltier effect can be expected by moving domain walls
\cite{Kovalev09,Bauer10} and mechanical rotations \cite{Bauer10}.

\section{Spin Seebeck effect \label{SSE}}

The most spectacular development in recent in years in the field of spin
caloritronics has been the discovery of the spin Seebeck effect, first in
metals \cite{Uchida08}, and later in electrically insulating Yttrium Iron
Garnet (YIG) \cite{Uchida10} and ferromagnetic semiconductors (GaMnAs)
\cite{Jaworski10,Bosu}. The spin Seebeck effect stands for the electromotive
force generated by a ferromagnet with a temperature bias over a strip of metal
normal to the heat current. This effect is interpreted in terms of a thermally
induced spin current injected into the normal metal that is transformed into a
measurable voltage by the inverse spin Hall effect
\cite{Saitoh06,Valenzuela,Kimura} metals. A separate Chapter of this book is
devoted to the spin Seebeck effect, so the present section is kept brief.

It is important to point out the difference between the spin Seebeck effect
and the magneto- or spin-dependent Seebeck effect measured by Slachter
\textit{et al.} \cite{Slachter} (see Sec. \ref{SDSE}). Both are generated at
an interface between a ferromagnet and a metal. In the magneto-Seebeck effect
a temperature gradient is intentionally applied over an intermetallic
interface, which is quite different from the spin Seebeck effect, and it can
be explained by traditional spin caloritronics concepts Johnson and Silsbee
\cite{Johnson87}. On the other hand, in the spin Seebeck effect the ISHE
contact is thermally floating and a standard thermoelectric explanation fails
\cite{Hatami10} (see, however, \cite{Nunner}).

There is consensus by now that the origin of the spin Seebeck effect is a net
spin pumping current over the ferromagnet/metal interface induced by a
non-equilibrium magnon distribution \cite{Xiao10,Adachi1}. Furthermore, the
phonon-magnon drag has been found to be very important
\cite{Adachi2,Jaworski11,Uchida11}. In magnetic insulators conventional
thermoelectrics cannot be applied. A longitudinal configuration in which a
temperature gradient is intentionally applied over the interface
\cite{Uchida10b} can therefore be classified a spin Seebeck effect. The
Slachter experiments \cite{Slachter} might also be affected by the spin
Seebeck effect, although the effect is probably overwhelmed by the
spin-dependent thermoelectrics.

As mentioned in Sec. \ref{SV}, the physics of the thermal torque induced by
heat currents in spin valves with an insulator as polarizing magnet as
proposed by Slonczewski \cite{Slonc} is identical to the longitudinal spin
Seebeck effect \cite{Uchida10b}, as explained theoretically by Xiao \textit{et
al}. \cite{Xiao10}. The \textquotedblleft loose\textquotedblright\ magnetic
monolayer model hypothesized by Slonczewski appears to mimic the solution of
the Landau-Lifshitz-Gilbert equation, which predicts a thin magnetically
coherent layer that effectively contributes to the spin pumping. Slonczewski's
claim that the heat current-induced spin transfer torque through magnetic
insulators should be large has been confirmed by first-principles calculations
that predict that the spin-mixing conductance at the interface between YIG and
silver is close to the intermetallic value \cite{Jia2}. This results is in
stark contrast to the expectations from a Stoner model for the magnetic
insulator \cite{Xiao10}, but can be explained by local magnetic moments at the
interface \cite{Jia2}.

From the discussion of the Onsager relations one might expect a spin Peltier
effect. To date no reports have been published on this topic, however.

\section{Conclusions}

The field of spin caloritronics has gained momentum in recent years since
experimental and theoretical groups have newly joined the community in the
last few years. It should be obvious from the above summary that many effects
predicted by theory have not yet been observed. The smallness of some effects
are also a concern. If spin caloritronics is to become more than a scientific
curiosity, the effects should be large enough to become useful. Therefore more
materials research and device engineering, experimental and theoretical, is
very welcome.

\acknowledgements

I am grateful for the most pleasant collaboration on spin caloritronics with
Arne Brataas, Xingtao Jia, Moosa Hatami, Tero Heikill\"{a}, Paul Kelly,
Sadamichi Maekawa, Eiji Saitoh, Saburo Takahashi, Koki Takanashi, Yaroslav
Tserkovnyak, Ken-ichi Uchida, Ke Xia, Jiang Xiao, and many others. This work
was supported in part by the FOM Foundation, EU-ICT-7 \textquotedblleft
MACALO\textquotedblright, and DFG Priority Programme 1538 \textquotedblleft
Spin-Caloric Transport\textquotedblright.

\end{document}